\documentclass[12pt,journal,draftcls,letterpaper,onecolumn]{IEEEtran}
\IEEEoverridecommandlockouts
\usepackage{cite}
\usepackage{amsmath,amssymb,amsfonts}
\usepackage{algorithmic}
\usepackage{graphicx}
\usepackage{textcomp}

\usepackage{xcolor}
\def\BibTeX{{\rm B\kern-.05em{\sc i\kern-.025em b}\kern-.08em
    T\kern-.1667em\lower.7ex\hbox{E}\kern-.125emX}}

\begin{document}

\title{Deep Learning Based Power Allocation Schemes in NOMA Systems: A Review\\}

\author{\IEEEauthorblockN{Zeyad Elsaraf}
\IEEEauthorblockA{\textit{School of Computing and Engineering} \\
\textit{University of Huddersfield}\\
Huddersfield, United Kingdom \\
zeyad.elsaraf@hud.ac.uk}\\
\and
\IEEEauthorblockN{Faheem A. Khan}
\IEEEauthorblockA{\textit{School of Computing and Engineering} \\
\textit{University of Huddersfield}\\
Huddersfield, United Kingdom \\
f.khan@hud.ac.uk}\\
\and
\IEEEauthorblockN{Qasim Zeeshan Ahmed}
\IEEEauthorblockA{\textit{School of Computing and Engineering} \\
\textit{University of Huddersfield}\\
Huddersfield, United Kingdom  \\
q.ahmed@hud.ac.uk}

}

\maketitle

\begin{abstract}
Achieving significant performance gains both in terms of system throughput and massive connectivity, non-orthogonal multiple access (NOMA) has been considered as very promising candidate for the future wireless communications technologies. It has already received serious consideration for implementation in the fifth generation (5G) and beyond wireless communication systems. This is mainly due to NOMA allowing more than one user to utilise one transmission resource simultaneously at the transmitter side and successive interference cancellation (SIC) at the receiver side. However, in order to take advantage of the benefits NOMA provides in an optimal manner, power allocation needs to be considered to maximise the system throughput.This problem is non-deterministic polynomial-time (NP)-hard which is mainly why the use of deep learning techniques for power allocation is required. In this paper, a state-of-the-art review on cutting edge solutions to the power allocation optimisation problem using deep learning is provided. It is shown that the use of deep learning techniques to obtain effective solutions to the power allocation problem in NOMA  is paramount for the future of NOMA based wireless communication systems. Furthermore, several possible research directions based on the the use of deep learning in NOMA systems are presented. 
\end{abstract}

\begin{IEEEkeywords}
Deep learning, Power allocation, Multiple access, Non-orthogonal multiple access (NOMA), 5G
\end{IEEEkeywords}

\section{Introduction}
Recently, the extraordinary growth of wireless technologies has led to  escalating demands for much greater data rates and increased capacity for emerging new applications such as augmented/virtual reality, Internet of Things (IoT) etc. To address these aforementioned challenges, several new technologies such as massive MIMO (multiple-input multiple output), millimeter wave (mmWave), and dynamic spectrum sharing (DSA) have been employed in the fifth generation (5G) wireless communication systems [1]-[7]. More efficient multiple access (MA) techniques have recently been proposed in 5G and beyond networks to achieve much higher system throughput and massive connectivity. One noteworthy MA technique that raises the system throughput while simultaneously increasing the connection density of the network is non-orthogonal multiple access (NOMA). As opposed to conventional MA schemes, in which users are assigned orthogonal resources such as frequency, code, time to circumvent inter-user interference, NOMA allocates these same transmission resources to all users in a non-orthogonal manner [8],[9].

This allows for users to be stacked in a non-orthogonal fashion which, in turn, allows for more than one user to occupy transmission resources at the same time. At the transmitter side, power allocation takes place according to the NOMA principle. That is, users with better channel conditions are allocated less power while users with worse channel conditions are allocated higher power. The signals of all users in the network are then superimposed upon each other, creating one superposed signal that is then broadcast through the channel. At the receiver side, users with worse channel conditions and thereby higher power levels detect their signals by treating all other user signals as noise, while users with lower power levels would need to carry out successive interference cancellation (SIC), as shown in Fig.1, to detect their signals. SIC aims to extract a specific user signal from the received superposed signal. It does this by initially decoding the signal(s) of the users with higher power levels, subtracting it from the superposed signal, and then decoding the difference as the user with a lower power level [10],[11]. As a result of the transmission and reception processes, a NOMA based system model has the ability to propel 5G communication systems towards achieving massive connectivity and high data rates. 

One key issue in NOMA system is to expertly carry out power allocation and joint channel assignment using limited transmission resources. Carrying out such processes has proven to be non-deterministic polynomial-time (NP)-hard [12], meaning, to find the optimum result, all feasible combinations of channel assignment need to be investigated, which has proven to be very computational complex as well as being quite expensive and thereby infeasible. That is primarily why researchers have put forth many heuristic and suboptimal approaches [12],[13] to solve this optimisation problem. Simultaneously, however, the solution space to the optimisation issue is quite large, making utilisation of nonlinear searching processes inevitable. 

\begin{figure}[h!]
   \begin{center}
    \includegraphics[width=0.85\linewidth]{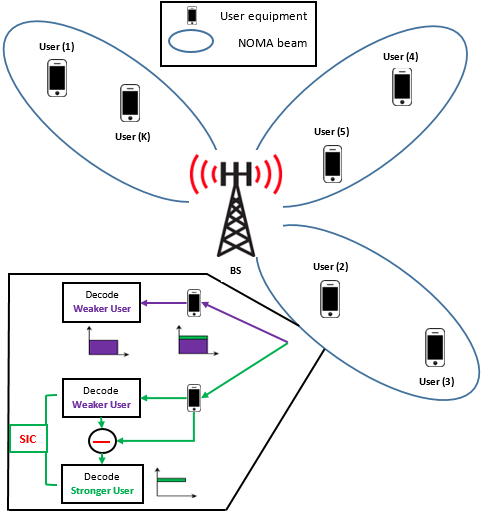}
    \caption{Power domain NOMA system model with multiple users}
    \label{fig:1}
		\end{center}
\end{figure}

That is why traditional methods of solving this optimisation problem are insufficient and unreliable in acquiring sufficient channel assignment and, by extension, making the performance of the NOMA system limited. Recently, deep learning has risen to prominence as an effective technique to be implemented in wireless communication systems, thereby improving their system design as Fig.2 [14] shows deep learning's general framework. By optimally utilising the non-linear relationship in the training data, deep learning enhances the system performance. Motivated by the power of deep learning, this paper provides a detailed overview of the role deep learning plays in power allocation in NOMA systems. 

The remainder of this work is divided into three sections. Section II presents a state-of-the-art review of power allocation in NOMA systems using deep learning techniques. The research conducted for this section focuses on very recent works on the topic, making this review on the cutting edge of upcoming technologies. Section III presents a number of challenging problems for power allocation using deep learning and a conclusion is drawn in section IV.

\section{Power Allocation using Deep Learning}

In order to most optimally utilize the advantages of the NOMA system, one very important issue is that of power allocation with limited resources. This problem of optimal power allocation is proved to be NP-hard, meaning, in order to acquire a solution that is optimal, all combinations of channel assignment must be investigated, which is not feasible if not very computationally complex and thereby expensive. Consequently, many solutions have been proposed by researchers to address this problem. Solutions such as, power allocation for a downlink single-input single-output (SISO) NOMA system, as shown in Fig.3 [15], power allocation for optimal user fairness [16], and power allocation for maximum energy efficiency [17]. Many of these solutions, however, were shown to be sub-optimal which is why the implementation of deep learning techniques is required. In the rest of this section, a comprehensive literature review on solutions to the power allocation problem using deep learning will be presented in adequate detail. As can be observed from Table 1, the reviewed works are on the cutting edge of recent technological advances in the field of power allocation using deep learning in NOMA. 

\begin{figure}[]
   \begin{center}
    \includegraphics[width=0.85\linewidth]{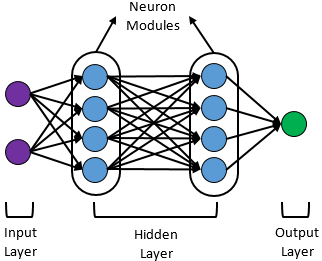}
    \caption{Deep neural network general framework}
    \label{fig:1}
		\end{center}
\end{figure}

\begin{table*}[]
\centering
\caption{\centering Deep learning based power allocation schemes in NOMA systems}
\label{tab:my-table}
\begin{tabular}{|c|c|l|}
\hline
\multicolumn{1}{|l|}{\textbf{Deep Learning Method}} & \textbf{Model}                                  & \textbf{Technical Contributions}                                                                                                                                                                                                                    \\ \hline
Reinforcement                                       & Q-learning                                      & \begin{tabular}[c]{@{}l@{}}base station (BS) performs power allocation with the \\ existence of an intelligent jamming device. The process is developed \\ as a game of the zero-sum variety. {[}20{]}\end{tabular}                                                               \\ \hline
Unsupervised                                        & K-means                                         & \begin{tabular}[c]{@{}l@{}}Examine the sum-rate maximization problem with \\ limited total transmission power while meeting \\ the users’ QoS. {[}17{]}\end{tabular}                                                                                \\ \hline
Supervised                                          & DNN  & \begin{tabular}[c]{@{}l@{}}Propose efficient up-to-date schemes for deep \\ learning (DL) based 5G and beyond scenarios. {[}18{]}\end{tabular}                                                                                                      \\ \hline
Reinforcement                                       & ANN & \begin{tabular}[c]{@{}l@{}}Propose an algorithm to optimally allocate transmission \\ resources by performing channel assignment through \\ an attention-based neural network (ANN) {[}15{]}\end{tabular}                                           \\ \hline
Supervised                                          & DNN                                             & \begin{tabular}[c]{@{}l@{}}In the presence of imperfect SIC, propose a power \\ allocation method with the aim of maximizing system \\ sum rate for a downlink NOMA scenario. {[}16{]}\end{tabular}                                                 \\ \hline
Federated                                           & DNN                                             & \begin{tabular}[c]{@{}l@{}}Propose a new scheduling policy as well as a novel \\ power allocation technique with the aim of maximising \\ the weighted sum data rate under practical constraints. \\ {[}19{]}\end{tabular}                          \\ \hline
Semi-supervised                                     & DNN                                             & \begin{tabular}[c]{@{}l@{}}Propose a DL framework to handle power allocation, \\ user association, and subchannel allocation while \\ maximising the energy efficiency (EE) of the system \\ while having a power limitation. {[}23{]}\end{tabular} \\ \hline
\end{tabular}
\end{table*}

[18] proposes a deep reinforcement learning (DRL) scheme to allocate power to users in optimal fashion, more precisely, ANN, is used to carry out channel assignment. The system model consists of one base station (BS) and multiple users in a downlink NOMA scenario. Deep learning  algorithm for reinforcement learning operates by treating BS as agent and users as performance environment. First, BS selects an action (channel assignment) from a set to assign channels and resources to users. Then, based on the reaction in the environment (the users), a feedback signal is sent back to the BS to contribute to the assigning of users in the next transmission. 

This process has three critical components, the state space, action space, and reward function. The state space is attributed to the channel information, which is represented as user channel pairs. The action space is where the agent (BS) takes an action that selects one channel for transmission of data for one user. To fulfil the needs of user channel allocation, the set of actions is limited; meaning each user corresponds to a different action. The allocation process is finished after N actions. The reward function is the signal fed back to the agent in response to a successful or unsuccessful transmission at each time slot end. The signal is comprised of the data rates experienced by each user transmitted to by the BS. 

The objective of [19] is to maximise this reward signal, thereby optimising the data rates experienced at each user. The results obtained show a sum-rate comparison between Joint Resource Allocation (JRA) with downlink (DL) and JRA without DL, with the DL version outperforming its non-DL counterpart by a significant margin. The results also contain spectral efficiency, minimum data rates, and sum rates for varying batch sizes and learning rates, with a batch size of 40 and learning rate of 0.001 shown to have the best performance. The DL based proposed algorithm is shown to outperform its non-DL counterparts on all different tests.

[20] proposes a power allocation scheme aiming to maximise system sum-rate for a downlink NOMA scenario with imperfect SIC using DL techniques. An exhaustive search algorithm is used for optimum power allocation. In this work, a power allocation method designed to maximise the experienced system sum rate is proposed in the presence of imperfect SIC. Through exhaustive search, the proposed method utilises deep learning to foretell the optimal power allocation factors. 

The system model consists of one BS that serves K users equipped with single antennas. The BS is placed at the centre of a cell and users are randomly placed within that cell. The proposed deep neural network (DNN) inputs such as: channel response normalised by noise (CRNN), total transmission power, and the fraction of signal power without successive interference cancellation. The inputs are all taken into consideration within the DNN with the aim being optimising the output which is a set of power allocation factors. The results obtained consist of a sum-rate versus total transmission power comparison for two values of uncancelled signal power for 2 users and for 3 users as well as the average central processing unit (CPU) processing time versus total transmission power for 2 users and 3 users. 

The simulation results showcase the designed technique acquires near optimal performance for the 2-user case and maintains this level of performance while suffering a decrease in performance in the low total transmission power range. The proposed scheme is also shown to be easy to run as its CPU processing time starts out low and remains constant over an increasing total transmission power while the optimal non-DL scheme’s processing time increases exponentially with an increasing total transmission power. The simulation results also show the suggested method can achieve a sum rate performance close to optimal levels while operating successfully with much lower computation complexity.

[21] studies the non-convex and hard to solve problem that is optimising power allocation for an increasing number of users for a mmWave NOMA system. This work examines the sum rate maximisation issue of the systems operating under the mm-wave NOMA protocol with the constraints of available total power and each user's required quality of service conditions. The developed optimisation issue is in nature a nonlinear issue and, by extension, is non-convex in nature and thereby very difficult to find a solution for, more so for a high number of users. The main driving motivation coming from the association characteristics of the channels of all users in the network in the mmWave NOMA network, a protocol based upon online k-means deep learning is developed for use in user grouping. Furthermore, an online K-means-based deep learning algorithm is proposed for application in realistic dynamic environments where new users continue to show up in a cumulative manner. 

The algorithm is also proposed to lessen the over-all computational complexity of the system. Moreover, to further improve the performance of the proposed algorithm, optimal power allocation policy is derived by making expert use of the successive decoding feature. The system model involves one BS and K users. The BS would broadcast separate beams containing the superposed varying messages of all the different users. Users would be grouped into clusters and served with NOMA beams, with beamforming taking place at the BS. The decoding order for SIC as per the NOMA principle is optimised according to the cluster and beamforming vectors. When both said vectors are fixed, the power allocation vector is known, which is then used to develop optimal power splitting factors for different clusters as well as for each user within each cluster.

\begin{figure}[]
   \begin{center}
    \includegraphics[width=0.85\linewidth]{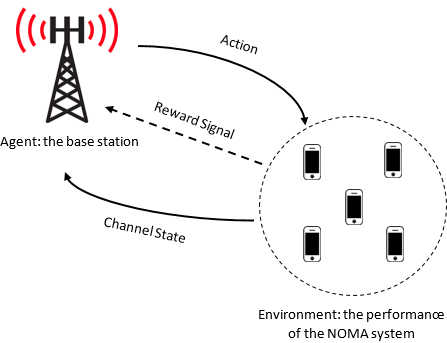}
    \caption{Power allocation and channel assignment using deep learning general system model [15]}
    \label{fig:2}
		\end{center}
\end{figure}

The simulation results obtained in this work contain two significant revelations. First, the proposed deep learning framework improves the functioning of systems operating with mmWave NOMA in comparison to traditional user grouping protocols such as complete re-clustering. The online k-means algorithm is also simulated using orthogonal multiple access (OMA) and compared with its NOMA counterpart where it is shown that the NOMA version has the superior performance. The user sum rate is evaluated through different types of user clustering, imperfect channel state information (CSI) environment, and compared with exhaustive search. The results show that the proposed K-means-based user clustering is superior in all tests. Secondly, the user grouping technique, which is based upon online k-means is shown to have similar functioning to traditional k-means while achieving an acceptable trade-off between experienced user sum rate and computational complexity.

The studies conducted in [22] suggest an optimal power allocation algorithm based on user ordering for SIC in a satellite Internet of things (IoT) NOMA system. Three algorithms are proposed in this work. For the first algorithm and based on a derived equation in the paper, each user in the system is assigned a value. Users are then ordered in the network in an ascending order based on said value. For the second algorithm, a deep learning-based power allocation is proposed, where a neural network is used as an approximation function deployed to ascertain the SIC decoding order in the chain based on a given queue state and channel state. 

The third algorithm proposes to take all different combinations for queue state and channel state and use it to train the deep neural network to produce an optimal, or near optimal, SIC decoding order. The system model consists of one satellite source node and K user terminals (terrestrial terminals) in a downlink scenario. Users are served via NOMA beams by sharing the frequency allocate to said beam. The aim of the proposed algorithms is to provide optimal power allocation via user ordering in the SIC decoding chain while avoiding channel congestion caused by a too high arriving rate or a too small leaving rate. The results show the third algorithm achieves the best performance in terms of data rates, mean square error, and queue delay.

[23] proposes a federated learning (deep learning) aided NOMA-based design that attempts to maximise the system sum rate. Federated learning (FL) is an extremely sought-after deep learning scheme that trains a model in a central manner as well as maintaining data dispersion. Distributed computation allows FL to be alluring for applications with a bandwidth constraint in wireless communication systems. There can be several distributed users connected to a central parameter server (PS) to download or upload data to or from the PS in an iterative manner. However, due to the constrained bandwidth, only a small part of the total number of devices in a connected state can be anticipated for each round. Due to the existence of numerous parameters in the state-of-the-art deep learning models, the system suffers from very complex computation as well as exerting a high communication load on dispersing and/or gathering data for training purposes. 

The study utilises graph theory to solve the optimisation problem for power allocation. The study also utilises the application of NOMA to increase the system accuracy while reducing the communication latency. The system model consists of one PS and M number of users in a downlink/uplink scenario. During the downlink phase, the PS initialises the model and broadcasts it to all users in the network. Each user then performs its local training task and outputs the gradient using local data. The output data is then sent back to the PS at the end of this phase upon which the PS recalculates the training model based on the received gradient data and re-broadcasts the result to all users again. This process would repeat itself until the model converges. 

The second phase, the uplink phase, is where NOMA is applied. Every user message would be detected at the PS as it applies SIC to decode user messages. Graph theory is applied at the second phase where it would be used to ascertain all possible user schedules then optimised power allocation would be performed to find the best one for better sum rate maximisation. The results obtained for this work show the testing accuracy for the proposed system model. The proposed optimal user scheduling and power allocation algorithm is shown to have the optimal performance when compared to its varying counterparts. The proposed scheme is also compared with conventional time division multiple access (TDMA), where it is shown to outperform it.

Communication systems are generally vulnerable to jamming attacks, more so in the case of NOMA on account of its inherent inter-user interference problem. In [24], a BS outfitted with numerous antennas is broadcasting against an intelligent jamming device. The power allocation for this work is developed as a game of the zero-sum variety. The leader, namely the BS, sets the transmit power on numerous antennas as the follower, namely the intelligent jammer, sets jamming signal power to interrupt users' broadcast process. Conditions guaranteeing the game’s functioning reveal the effects of numerous antennas and channel radio states. A technique regarding the control of power levels is allocated on DRL is presented for the downlink NOMA scenario, the transmission of which is carried out while simultaneously being unaware of the existence of the jamming signal and the channel parameters of the radio channel.

The system model is comprised of one BS, one smart jammer, and M users, all equipped with N antennas. A time-slotted NOMA system is considered with antennas to receive transmitted signals along with the interference of an intelligent jammer. BS utilises its transmit antennas to broadcast a signal that is superposed according to the NOMA framework while the smart jammer uses its antennas to broadcast an interference signal. The jamming signal power is chosen as attested by the ongoing downlink transmit power. The MIMO channel can be examined as an independent sub-channels group and hence the channel gain matrices of the users can be arranged by the squared Frobenius norm [25], [26]. The receiver noise as well as the gains of the channels are assumed to be in an independent state for all users and each other. The derivations made in this work show jamming resistance performance of the system rises along with channel power gains and the number of antennas. The proposed quick Q-based protocol for NOMA power assignment combining hot-booting and Dyna architecture is presented regarding a changing game to speed up learning and hence enhances system performance with the inclusion of smart jamming. 

Simulation results show that proposed NOMA protocol for power assignment is able to remarkably enhance utility and overall data rate as well as average signal-to-interference-plus-noise ratio (SINR) of BS very early after the Stackelberg game start. A possible future direction for this work is to possibly extend the theoretical and experimental findings to more realistic environments with smart jamming. An example of said environment is to have a jammer use programmable radio devices to choose multiple jamming policies in a flexible fashion.

The integration of unmanned aerial vehicles (UAVs) into visible light communication (VLC) has the potential to offer numerous advantages for applications requiring massive connectivity and services in 5G and beyond. [26] considers a UAV-assisted VLC using NOMA. More Precisely speaking, a joint problem of power allocation is formulated as well as UAV’s placement in order to maximise the sum rate for all users in the network, which are still subject to power allocation constraints, UAV’s position, and the quality of service expected at each user. Since the aforementioned problem in this work is non-convex in nature, it is difficult to solve in an optimal manner. Hence, this study proposes utilising the swarm algorithm Harris hawk’s optimisation (HHO) algorithm to attempt to find a solution to the presented problem and achieve a workable solution. 

The system model considered is comprised of one BS, namely the UAV, which is outfitted with a transmitter of the light emitted diode (LED) kind to supply communications as well as illumination for a number of placed ground users (GU), in a random manner, under its coverage area. The transmitter of the LED kind that is outfitted onto the UAV broadcasts data at the same time to the GUs using the identical transmission resource of time and frequency by adequately configuring the available power used for transmission. GUs have to apply SIC in order to extract their messages and signals by removing the other messages attached to the other GUs. According to the NOMA principle, the decoding order follows an increasing order of the channel gains. For the channel model, the line of sight (LoS) path of propagation is considered. Generalised Lambertian emission model is used where the DC gain is in proportion to the inverse of the squared distance between the LED transmitter and the user [27]. The simulated results and performance metrics show the efficacy of the proposed HHO algorithm. Moreover, the obvious gain of jointly optimising NOMA assignment of power levels and the positioning of the UAV can be observed in comparison to conventional techniques of optimising the positioning of the UAV or power allocation. 

\section{Future Research Directions}
In this section, possible future research directions will be listed and discussed. The following are examples for future work in the area of the use of deep learning in NOMA systems:
\begin{enumerate}
    \item \textbf{Dynamic mmWave NOMA:} Most of the above mentioned solutions for optimal power allocation do not consider more developed reinforcement and online learning processes which update the partition according to dynamic mmWave NOMA scenarios, that is a possible avenue for future work.
    \item \textbf{Smart Jamming:} Some of the reviewed works can be extended to include more practical applications for NOMA transmissions, specifically with the presence of smart jamming, where a programmable jammer makes use of radio devices to choose jamming policies in a flexible manner.
    \item \textbf{Multiple Antennas and Mobile Users:}  Some of the reviewed works can be extended to users equipped with multiple antennas as well as considering the case for mobile users, in which the power allocation and beam selection need to be updated in real time.
    \item \textbf{Data Set Acquisition:} The quantity and quality of the testing data have a noticeable effect on the performance of the system based on deep learning. With the rise of natural language processing (NLP) and computer vision, there exist data sets that can provide adequate performance gains for training the system. However, there need to be an improved data set that can provide reliable and robust solutions for the many problems facing NOMA systems including power allocation. 
    \item \textbf{Model Selection:} Considering deep learning based communication systems, designing the neural network is the main problem. Given deep learning's potential to not only streamline and simplify the communication process but also reduce the complexity and thereby the cost, it becomes imperative to find general frameworks such as long short-term memory (LSTM), which is widely used for NLP. There exists the need for effective and general models that can address the issue of power allocation in NOMA.
    \item \textbf{Deep Reinforcement Learning in NOMA:} Deep reinforcement learning (DRL) is an alternative method for solving resource allocation issues. For the next generation of wireless communication technologies, there exist numerous energy management and resource allocation optimisation problems, including power allocation issues, but the current level of deep learning knowledge is ineffective at handling large amounts of data. DRL is a promising candidate for handling this issue which is why it needs to be investigated thoroughly for future wireless communications using deep learning models. 
    \end{enumerate}
There are, however, numerous other possible future research directions including, but not limited to, learning mechanism and performance analysis and model compression for deep learning based 5G and beyond systems.

\section{Conclusion}
In this paper, a thorough review of power allocation using deep learning in NOMA systems has been presented. The topics covered include, power allocation for sum rate maximisation with imperfect SIC, unsupervised deep learning for mm-wave NOMA systems, and power allocation in a NOMA system based on federated learning among others. A number of future research directions have also been discussed such as, dynamic mm-wave NOMA, NOMA with smart jamming, and NOMA with multiple antennas and mobile users. This survey paper has shown, without a doubt, that the implementation of deep learning in NOMA is paramount to advance the performance of NOMA based wireless communication systems in order to address the challenges posed by 5G and beyond systems.

\end{document}